# Modeling microbial cross-feeding at intermediate scale portrays community dynamics and species coexistence

Chen Liao[1], Tong Wang[2,3], Sergei Maslov[3,4], Joao B. Xavier[1,*]

[1]Program for Computational and Systems Biology, Memorial Sloan-Kettering Cancer Center, New York, NY, United States of America, [2]Department of Physics, University of Illinois at Urbana-Champaign, IL 61801, USA, [3]Carl R. Woese Institute for Genomic Biology, University of Illinois at Urbana-Champaign, IL 61801, USA, [4]Department of Bioengineering, University of Illinois at Urbana-Champaign, IL 61801, USA

[*]To whom correspondence may be addressed: J. Xavier (Email: XavierJ@mskcc.org)

Keywords:

cross-feeding; coarse-grained modeling; microbial community; ecological relationship; species coexistence; population dynamics

# Abstract

Social interaction between microbes can be described at many levels of details, ranging from the biochemistry of cell-cell interactions to the ecological dynamics of populations. Choosing the best level to model microbial communities without losing generality remains a challenge. Here we propose to model cross-feeding interactions at an intermediate level between genome-scale metabolic models of individual species and consumer-resource models of ecosystems, which is suitable to empirical data. We applied our method to three published examples of multi-strain *Escherichia coli* communities with increasing complexity consisting of uni-, bi-, and multi-directional cross-feeding of either substitutable metabolic byproducts or essential nutrients. The intermediate-scale model accurately described empirical data and could quantify exchange rates elusive by other means, such as the byproduct secretions, even for a complex community of 14 amino acid auxotrophs. We used the three models to study each community's limits of robustness to perturbations such as variations in resource supply, antibiotic treatments and invasion by other "cheaters" species. Our analysis provides a foundation to quantify cross-feeding interactions from experimental data, and highlights the importance of metabolic exchanges in the dynamics and stability of microbial communities.

## Significance statement

The behavior of complex multispecies communities such as the human microbiome is hard to predict by its composition alone. Our efforts to engineer such communities would benefit from mechanistic models that accurately describe how microbes exchange metabolites with each other and how their environment shapes these exchanges. But what is the most appropriate level of details to model microbial interaction? We propose an intermediate level to model metabolic exchanges that accurately describes population dynamics and stability of microbial communities. We demonstrate this approach by constraining models with experimental data from three laboratory communities with increasing levels of complexity. Each model allows us to predict metabolic byproduct leakage fractions as well as how external perturbations such as nutrient variations or addition of antibiotics impact those communities. Our work paves the way to model real-world applications including precise engineering of the microbiome to improve human health.

# Introduction

Most microorganisms that affect the environments we live in[1] and that impact our health[2] do not live in isolation: they live in complex communities where they interact with other strains and species. The past decade has seen a surge of scientific interest in microbial communities, such as the human microbiome, but most studies remain limited to cataloguing community composition[3]. Our mechanistic understanding of how biochemical processes occurring inside individual microbial cells command the interactions occurring between cells, and lead to the emergent properties of multi-species communities remains limited[4].

Microorganisms consume, transform and secrete many kinds of chemicals, including nutrients, metabolic waste products, extracellular enzymes, antibiotics and cell-cell signaling molecules such as quorum sensing autoinducers[5,6]. The chemicals produced by one microbe can impact the behaviors of other microbes by promoting or inhibiting their growth[7], creating multi-directional feedbacks that drive ecological interactions which may be beneficial or detrimental to the partners involved[8,9].

If a community is well-characterized and given sufficient data on population dynamics, it should be possible to parameterize the underlying metabolic processes involved in microbe-microbe interactions by fitting mathematical models[10]. Any model can potentially yield insights[11], but the complexity of most models so far has been either too high for parameterization, or too low to shed light on cellular mechanisms. Microbial processes may be modelled across a range of details: At the low end of this spectrum of details we have population dynamic models such as generalized Lotka-Volterra (gLV)[12] and Consumer-Resource (C-R) models[13], which treat each organism as a 'black-box' at the cellular level. For example, C-R models assume a linear or Monod dependence of microbial growth on resource uptake kinetics. At the high end of this spectrum, we have detailed single-cell models such as dynamic flux balance analysis (dFBA)[14] and agent-based models[15] that have too many parameters to be parameterizable by experimental data. For example, the linear equations for fluxes obtained from quasi-steady-state assumption of dFBA are highly underdetermined. What is the appropriate level of details to model and constrain microbial processes using data, that may produce not only accurate predictions but also mechanistic insights on microbial communities?

Here we propose a generalizable framework that couples classical ecological models of population and resource dynamics with coarse-grained intra-species metabolic networks. We show that modeling communities at this intermediate scale can accurately quantify metabolic processes from population dynamics data alone. We demonstrate the value of this approach on three engineered communities of *Escherichia coli* (*E. coli*) strains with increasing levels of complexity: (1) unilateral acetate-mediated cross-feeding[16], (2) bilateral amino-acid-mediated cross-feeding between leucine and lysine auxotrophs[17], and (3) multilateral amino-acid-mediated cross-feeding between 14 distinct amino acid autotrophs[18]. The models report inferred leakage fractions of metabolic byproducts that are difficult to measure directly by experiments, reveal how resource supply and partitioning alter the coexistence and ecological relationships between cross-feeders, and predict the limits of community robustness against external perturbations.

## Results

**Modeling microbial metabolic processes at an intermediate level is appropriate to fit the population dynamics data.** Inspired by the classical MacArthur's CR models[19] and many follow-ups[13,20–22], we propose to integrate CR models with a coarse-grained yet mechanistic description of cell metabolism. Metabolic reactions can be broadly classified as catabolic and anabolic, where

catabolic reactions break down complex substrates from culture media into smaller metabolic intermediates that can be used to build up biomass components by anabolic reactions. A minimal representation of cell metabolism is a three-layer network composed of growth substrates at the top, metabolic intermediates in the middle, and biomass at the bottom (Fig. 1). Despite its simplicity, this model is flexible enough to describe the transformation of resources into other resources or non-consumable chemicals and biomass, regardless of the specific reactions involved. Real cells can consume multiple nutritional resources that may be either substitutable or complementary for cell growth. Our model focuses on complementary resources for three reasons: (1) many microorganisms in natural samples are auxotrophs[23] whose growth relies on complementary essential nutrients; (2) minimal medium—popular for cultivating microbial communities in laboratory conditions including the data analyzed in our study—is composed of complementary nutrients; (3) substitutable metabolites can be mathematically lumped into functional groups.

Based on these assumptions, we developed a dynamic modeling framework that contains six kinds of biochemical reactions describing resource consumption, transformation, secretion, utilization for biomass synthesis, and inactivation (Supplementary Equations (S1)-(S6)). Briefly, substrates available in the growth media can be imported into cells. A certain fraction of the imported substrates is then broken down into metabolites, which can either be released back to the surrounding environment or used by cells for biomass production. Released metabolites can be imported by cells in a way similar to externally supplied substrates, except that their uptake may be inhibited by other substitutable substrates that are assumed to be preferentially used. To model the effects of toxic compounds[24] we allow the growth rate of any cell population to be not only governed by a birth-death process that constantly produces and loses cell material due to biosynthetic and maintenance processes respectively, but can be additionally inhibited by accumulation of toxic metabolites in the environment.

The six types of reactions can be translated to differential equations by specifying their kinetic rate expressions. We assumed quasi-steady-state for intracellular substrates and metabolites, as metabolic reactions typically occur at faster time scales compared to ecological dynamics. The time-scale separation thus simplifies our model by excluding intracellular variables, leaving only three types of variables that describe the population density of active cells ($N_l$, $l = 1,2,\cdots,n_c$), the extracellular concentrations of substrates ($[S_i]$, $i = 1,2,\cdots,n_s$), and the concentrations of metabolic byproducts excreted by cells ($[M_j]$, $j = 1,2,\cdots,n_m$). Assuming a chemostat environment with dilution rate $D$ (which reduces to a batch culture when $D = 0$), the differential equations associated with the three state variables are given below

$$\frac{d[S_i]}{dt} = D(S_{0,i} - [S_i]) - \sum_{l=1}^{n_c} J_{l,i}^{upt,S} N_l \qquad (1)$$

$$\frac{dN_l}{dt} = N_l(J_l^{grow} - J_l^{death} - D) \qquad (2)$$

$$\frac{d[M_j]}{dt} = D(M_{0,j} - [M_j]) + \sum_{l=1}^{n_c} (J_{l,j}^{leak,M} - J_{l,j}^{upt,M}) N_l \qquad (3)$$

where $S_{0,i}$ and $M_{0,j}$ are the feed medium concentrations of substrate $S_i$ and metabolite $M_j$ respectively. $J_{l,i}^{upt,S}$ and $J_{l,j}^{upt,M}$ represent uptake fluxes of substrates and metabolites respectively, $J_{l,j}^{leak,M}$ are metabolite secretion fluxes, and $J_l^{grow}$ and $J_l^{death}$ stand for per-capita growth and death rates respectively. We used Monod kinetics and first-order kinetics for resource uptake ($J_{l,i}^{upt,S}$ and $J_{l,j}^{upt,M}$) and cell death ($J_l^{death}$) respectively, and obtained expressions for resource transformation into other resources ($J_{l,j}^{leak,M}$) and biomass ($J_l^{grow}$) by intracellular flux balance analysis. The functional forms of these kinetic laws and other details of model formulation are described in Supplementary Texts 1.1.

Experimental data can be used to determine the parameters of our model either manually (by visual inspection) or automatically (by optimization algorithms). In the examples below we applied a combination of automatic and manual calibrations, where the latter is arguably a subjective process and requires an experienced operator with prior knowledge to choose a set of parameter values that are physically and biologically realistic through a laborious trial-and-error process. For each application, the manual process of parameter estimation began with initial values of parameters selected to be either equal to their previously reported values or assumed to be of the same order of magnitude based on the literature data. This was followed by the iterative evaluation of model outputs and refinement until sufficient concordance between the model predictions and the experimental data is achieved.

**Fitting the model to microbial community data.** We applied our framework to published datasets of two two-species communities with increasing level of complexity: a uni-lateral[16] and a bi-lateral[17] cross-feeding between laboratory evolved and engineered strains of *E. coli* respectively. Our goal was to manually parameterize the intrinsic metabolic processes relevant for the interactions between the community members, directly from time series data of community composition and experimentally measured metabolite concentrations. The number of metabolites essential for *E. coli* growth is estimated of the order of hundreds[25]. Therefore, we chose to include in our model as model variables only the metabolites known to mediate interpopulation interactions, together with the most limiting growth substrate.

The first community is a well-documented unilateral acetate-mediated cross-feeding polymorphism evolved from a single ancestral lineage of *E. coli* in laboratory conditions[16] (Fig. 2A, Supplementary Texts 1.2.1, and Supplementary Table 1). The community contains two polymorphic subpopulations (*E. coli* subspecies) whose metabolism differs in their quantitative ability to uptake and efflux carbon sources: a glucose specialist strain (CV103) which has a faster glucose uptake rate but cannot grow on acetate, and an acetate specialist strain (CV101) which can grow on acetate but has a lower glucose uptake rate. CV103 secretes acetate—a major by-product of its aerobic metabolism—and this way creates a new ecological niche for CV101. Fig. 2B-E shows that our model accurately reproduced the observed changes in growth and acetate concentration in both monoculture and coculture experiments over time. Particularly, we captured that the competition outcome depends on the acetate level in the feed medium (Fig. 2E), which can be explained by the positive nutritional effect of the acetate at low concentrations (Supplementary Fig. 1).

The second community is characterized by a synthetic cross-feeding mutualism between lysine and leucine auxotrophs of *E. coli*[17] (Fig. 2F, Supplementary Texts 1.3.1, and Supplementary Table 2). The two mutants differ only by single gene deletions in the lysine (Δ*lysA*) and leucine (Δ*leuA*) biosynthesis pathways. Neither mutant can grow in monoculture, but their coculture can survive by creating a bilateral dependency of two mutants cross-feeding each other missing essential amino acids. Fig. 2G, H show that our model was able to quantitatively recapitulate the growth and nutrient dynamics in both monoculture and coculture conditions. The fitted values of parameters reveal that the maximum growth rate of the lysine auxotroph is over 50% larger than that of the leucine auxotroph (Fig. 2I), which is consistent with the data showing that the biosynthesis of leucine is more costly than the biosynthesis of lysine[18]. Nonetheless, the parameters also indicate that the mortality rate of the lysine auxotroph (about 20% of its maximum growth rate) is also substantially higher than that of the leucine auxotroph (Fig. 2J), which qualitatively agrees with cell viability experiments in the monoculture and absence of amino acid supplementation[17]. Since cell mortality rate is determined by the ratio of maintenance rate to nutrient recycling efficiency from dead cells[26], this finding suggests that the lysine auxotroph has either or both of high maintenance cost and low biomass recovering yield.

Comparison of these two cross-feeding models suggests that resource sharing between natural (CV103 and CV101) and engineered (Δ*lysA* and Δ*leuA*) cross-feeders can be markedly different. We predicted that the glucose specialist lost 33% carbon in acetate overflow resulting in nearly equal flux values between acetate secretion and glucose uptake, a quantitative relationship that has been observed in a different *E. coli* strain[27]. By contrast, the engineered interaction between the Δ*lysA* and Δ*leuA* is much weaker with only 0.3% and 1.4% carbon loss in releasing leucine and lysine respectively. Although the acetate-mediated cross-feeding may have been an incidental finding, the high efflux of acetate could facilitate adaptive co-evolution and accumulation of degenerative mutations[16].

**Metabolic secretion fluxes modulate likelihood of genotypic coexistence.** The stable coexistence of different genotypes is a prerequisite for mixed microbial communities. But how strong are the metabolic secretion fluxes necessary to maintain genotypic coexistence in the absence of metabolite supplementation? We leveraged the two cross-feeding models above to address this question by simulating cocultures in chemostats at varied levels of resource supply and partitioning, which independently and synergistically modulate the actual secretion flux values.

We constructed phase diagrams that show how the community composition at steady state has distinct patterns between the two cross-feeding systems (Fig. 3A,B). First, competitive exclusion does not occur when cross-feeding is obligate and bidirectional (Fig. 3B). Second, coexistence of the glucose and acetate specialists can be attained largely independent of glucose supply when the partitioning level, controlled by the acetate leakage fraction $\varphi_a$, is below a certain threshold (dashed yellow line in Fig. 3A). By solving the model analytically (Supplementary Texts 1.2.2), we found that the threshold can be approximated by $\Delta V_g = (V_{3,g} - V_{1,g})/V_{3,g}$, where $V_{3,g}$ and $V_{1,g}$ are the maximum glucose uptake rates of the glucose and acetate specialists respectively. When $\varphi_a > \Delta V_g$, the glucose specialist releases more acetate than the amount needed to help the acetate specialist overcome its basal growth disadvantage, causing a declining self-balancing capacity of population dynamics and reduced likelihood of coexistence. By contrast, coexistence of the lysine and leucine auxotrophs is only weakly constrained by the resource partitioning level, but ultimately determined by the total amount of resources put into the system (Supplementary Texts 1.3.2).

Within the region of coexistence, the relative frequency of the acetate specialist increases continuously with the fraction of acetate leaked (Fig. 3A), whereas increasing the fraction of lysine leaked by the leucine auxotroph triggers a discontinuous, abrupt switch from a steady state dominated by the leucine auxotroph to a steady state dominated by the lysine auxotroph (Fig. 3B). Such abrupt, discontinuous regime shifts are a common feature of microbial communities limited by several essential nutrients[28]. Interestingly, growth of the dominant and rare auxotrophs are always limited by its auxotrophic amino acid and glucose respectively, which suggests an implicit negative feedback loop that maintains their relative abundance ratio before and after the switch: increasing population size of the dominant auxotroph impairs the growth of the rare auxotroph by consuming more glucose but eventually, its own growth is inhibited because a smaller amount of amino acid it needs to grow can be produced by its partner. Taken together, our models show that the likelihood of coexistence can be modulated by varying the metabolic secretion fluxes, but the effect of varying those fluxes depends on the approach used to modulate the system (resource supply or partitioning) and the cross-feeding type (unilateral or bilateral).

**Environmental changes to nutrients can reverse the sign of microbial social interactions.** Cross-feeding interactions within a microbial community may be described as social interactions with costs and benefits to the members involved[29,30]. Those costs and benefits may be altered by environmental perturbations that supply or remove the cross-fed metabolites form the environment. Using our community model, we investigated how the supplementation of metabolite mediators affected ecological relationships between cross-feeders at the steady state. We simulated chemostat cocultures at increasing levels of metabolite supplementation in the feed medium, and computed the net effect (+,0,-) of one population on the other by comparing to monoculture simulation. The pairwise ecological relationship between the two populations can then be determined by the signs of their reciprocal impacts[31].

The ecological relationship between the glucose and acetate specialists was displayed on a 2-dimensional phase space spanned by the feed medium concentrations of glucose and acetate (Fig. 4A). The entire space is divided into six distinct regions with diverse outcomes, including population collapse, competitive exclusion, and stable coexistence. Notably, it is very difficult to select supplementation resulting in stable coexistence. This is because, as explained above, the inferred value of $\varphi_a$ (0.33) is much greater than that of $\Delta V_g$ (0.12). The remaining diversity of the phase space structure is primarily driven by the dose-dependent effect of acetate[24]: it serves as a nutrient for the acetate specialist at low concentration but becomes inhibitory to growth of both strains when abundant (Supplementary Fig. 1). To illustrate this effect, we increased glucose supplementation from $P_1$ to $P_3$ (gray dots in Fig. 4A) in the phase space, which induced higher release of acetate to environment (Fig. 4B, top row) and switch of winners of the coculture competition (Fig. 4B, middle row). The glucose specialist wins the competition at $P_1$ because acetate level is too low to compensate the growth disadvantage of the acetate specialist. From $P_1$ to $P_2$, acetate concentration exceeds the threshold level of compensation and thus supports faster growth of the acetate specialist. Further increase of acetate concentration to $P_3$ inhibits both strains, among which the acetate specialist is more susceptible (Fig. 4B, bottom row; see also Fig. 2D): therefore, the glucose specialist wins again when the negative inhibitory effect of acetate outweighs its positive nutritional effect on the acetate specialist.

Compared to unilateral cross-feeding, new ecological relationships such as mutualism and parasitism emerges in the phase space when cross-feeding is bidirectional (Fig. 4C). The mutualistic relationship was maintained over a broad range of supplied amino acid concentrations, even though amino acid supplementation releases the dependence of one auxotroph on the other and is hence detrimental to mutualism. In the regime of mutualism, glucose is in excess and both strains are limited by the essential amino acids they cannot produce (Fig. 4D, left column). Further addition of amino acids leads to strain dominance, but not necessarily competitive exclusion. The lysine auxtroph was excluded when leucine was provided to release the leucine auxotroph from its growth dependence (Fig. 4D, middle column), whereas adding lysine only reduced the relative abundance of the leucine auxotroph, rather than leading to the loss of its entire population (Fig. 4D, right column).

Amino acid supplementation may lead to competitive exclusion or parasitism depending on whether one or both auxotrophs are limited by glucose. When glucose limits both auxotrophs, the leucine auxotroph wins because it has the same growth rate as the lysine auxotroph on glucose but lower death rate (Fig. 2I,J). When only the lysine auxotroph is limited by glucose, the leucine auxotroph can sustain its population by occupying a different niche and growing on leucine released by its competitor. Regardless of the outcome, our results suggest that adding cross-fed nutrients can induce competition between community members that previously interacted mutualistically, and shift positive interactions to negative interactions.

**Uncovering complex cross-feeding interactions between 14 amino acid auxotrophs.** Next we demonstrated the utility of our model to study cross-feeding interactions within communities of more than two members. We modeled a community of 14 amino acid auxotrophs engineered from *E. coli* by genetic knockout[18]. The 14-auxotroph model was directly extended from our 2-auxotroph model (Supplementary Texts 1.4.1) by considering each auxotroph can potentially release all other 13 amino acids to the shared environment. Although all feeding possibilities are known, the consumer feeding preferences are not. By fitting experimental data on the population compositions we aimed to infer the unknown feeding pattern—what amino acids and how much they are released by each auxotrophic strain to feed each other.

The model constructed this way has a total of 269 parameters; 50 of these parameters are either biological constants or can be obtained from the literature (Supplementary Table 3). From the remaining parameters, the 196 unknown amino acid leakage fractions (14 auxotroph by 14 amino acids) can be easily estimated by automatically minimizing the least square error between observed fold changes of population density in all pairwise batch cocultures (196 data points in total) and their analytical, rather than simulated, solutions after model simplication (Supplementary Texts 1.4.2).

Outcompeting a simple population dynamics model (Fig. 5A, Pearson's correlation coefficient = -36.06%), our fit gave an excellent match to the data (Fig. 5B, Pearson's correlation coefficient = 94.32%), except for cross-feeding pairs whose observed fold change values are less than 1. The observed reduction of growth fold changes may be caused by cell death in the absence of nutrients but practically, we assumed no cell death (so simulated growth fold changes are always non-decreasing) because measurement of optical density at low inoculation amount ($10^7$ cells/mL) is highly noisy and we are unable to distinguish between the two factors. Clearly, the 14 auxotrophs derived from the same wild-type strain showed different profiles of amino acid leakage (Fig. 5C): some auxotrophs such as the methionine auxotroph $\Delta M$ (36.41% total carbon loss) are highly cooperative whereas others such as the tryptophan auxotroph $\Delta W$ (1.37% total carbon loss) have very low cooperativity.

The remaining 20 free parameters, among which 14 are death rate constants, were obtained by manually selecting a set of values that fit the population dynamics of serially diluted cocultures of all 14 auxotrophs and four selected 13-auxotroph combinations (Fig. 5D). The fit is reasonably good at the log scale, except for the $\Delta M$-absent community which seems to undergo non-ecological processes that rescue the threonine auxotroph ($\Delta T$) from the brink of extinction between day 2 and day 3. Quantitatively, the Pearson's correlation coefficients between log10-transformed observed and predicted values are 88.71% (all 14 auxotrophs), 75.30% ($\Delta K$-absent), 78.34% ($\Delta R$-absent), 52.93% ($\Delta T$-absent), and 8.90% ($\Delta M$-absent). Most auxotrophs were diluted away very quickly but some exhibited transient recovery dynamics after the initial decay. For example, population density of the isoleucine ($\Delta I$) auxotroph had an initial drop because the isoleucine pool had not been accumulated to a critical size that allows the actual growth to compensate for mortality and dilution. As the pool size increases, its net growth rate (growth minus mortality) surpasses the dilution rate and recovers its population density, which eventually levels off when the positive and negative forces reach equilibrium. By fitting the population density dynamics, we concomitantly inferred the concentration dynamics of glucose and all amino acids (Supplementary Fig. 3), which are hidden states (not yet observed) that are relatively costly and inaccurate to measure in experiments.

**Cross-feeding network is prone to collapse upon external perturbations.** By simulating the 14-auxotroph community model to steady state, we predicted that the initial mixture converges to a stable coexisting subset that contains 4 auxotrophs that are deficient in biosynthesis of isoleucine ($\Delta$I), lysine ($\Delta$K), methionine ($\Delta$M), and threonine ($\Delta$T) (Fig. 6A). The predicted coexistence state was successfully validated by two independent observations over 50-day serial dilution[18], a much longer period of time than the duration of the training dataset (7-day serial dilution; Fig. 5D). The predicted resource-consumer relationships of the stable subset are shown in a bipartite network (Fig. 6B), where 3 amino acid secretion fluxes were identified as essential (solid arrows) as their deletions resulted in strain loss (Supplementary Fig. 4). These essential fluxes suggest that the primary feeders for $\Delta$K, $\Delta$M, $\Delta$T are $\Delta$T, $\Delta$I, $\Delta$M respectively; however, none of $\Delta$K, $\Delta$M, $\Delta$T dominates the feeding of $\Delta$I and their contributions to the isoleucine pool in the environment are substitutable.

We computationally tested how external perturbations, including nutrient downshift, the addition of antibiotics, and invasion of cheating phenotypes (the same auxotrophic dependence but no amino acid leakage) affect the stability of coexistence among the 4 auxotrophs (see Methods).

The 4-strain community was able to cope with these disturbances to a certain extent and remained integrated. Beyond the thresholds, all three perturbation types resulted in community collapse as a result of domino effect (Fig. 6C-E), implying that tightly coupled cooperative communities are fragile and prone to collapse. Since antibiotics inhibit growth of individual strains (targeting consumer nodes in the bipartite network) while cheaters are amino acid sinks (targeting resource nodes in the bipartite network), we identified that ΔT and methionine as the weakest consumer node (Fig. 6D) and resource node (Fig. 6E) in the bipartite network respectively. Our results suggest that ΔT→K (secretion of lysine by the threonine auxotroph) and M→ΔM (uptake of methionine by the methionine auxotroph)—the outgoing links from the two weakest nodes that are also essential to maintain community integrity—are the weakest metabolic fluxes that may set the resistance level of the community to external perturbations[32].

## Discussion

Predicting population dynamics from the interactions between its members is difficult because interactions can happen across multiple scales of biological organization[33]. Here we propose a coarse-grained yet mechanistic ecology model and show that it may accurately quantify the metabolic exchanges underlying cross-feeding interactions in well-defined laboratory communities. Previous studies have used the metabolic flux analysis, but these studies required flux measurements by isotope tracing and metabolomics to fit the adjustable flux parameters in a stoichiometric metabolic model. Some success was also achieved by fitting the time series data with simple ecological models[34–38] such as the gLV equations; however, in gLV-type models, interspecific interactions are phenomenologically defined based on density dependency, which gives no mechanistic understanding of how interactions occur[39]. By contrast, our model has explicit formulations of context dependency by representing the chemical flows within and between microbes and thus can explain the metabolic part of microbe-microbe interactions.

When we have limited prior knowledge and data on a given community it becomes critical to choose the right level of details. We show that a highly detailed metabolic network is not necessary for developing useful ecological models. In single-bacteria studies, coarse-grained metabolic models have been employed to understand the design principles of metabolic networks and their regulation[40], as well as to predict metabolic flux distributions useful for synthetic biology[41] and industrial[42] applications. Compared to genome-scale models, using coarse-grained models linking ecology and metabolism is simple but rarely done until recently[22]. Depending on the research question, a coarse-grained metabolic network can be created at any level of granularity from a single reaction to the complete genome-scale reconstruction. The choice of granularity and how to derive a simpler model from the more complex one are usually empirical but can be facilitated by more systematic approaches to reduce dimensionality.

Our model could extract new insights from previously published empirical data. The analysis shows that unidirectional cross-feeding is equivalent to a commensalism and bidirectional cross-feeding is equivalent to a mutualism. As shown by our study (Fig. 4) and previous work[24,29], the actual relationship between cross-feeders, however, can be diverse in even simple and constant environments (e.g., glucose minimal medium) due to a combination of positive effects of cross-feeding with negative effects of competition and toxicity of cross-fed metabolites, suggesting that the exact outcome cannot be precisely delineated by the cross-feeding type alone. Moreover, mechanistic models can help identify knowledge gaps[43]. For example, recent experiments have demonstrated that the coexistence of two carbon source specialists in the unilateral cross-feeding example is mutualistic in the sense that the consortium is fitter than the individuals[44]. The syntropy can be explained by a null expectation from theoretical ecology models[45]: the glucose specialist provides acetate in an exchange for a service provided by the acetate specialist which scavenges the acetate down to a level at which growth inhibition is insignificant. Although we thoroughly considered the mechanism of resource-service exchange, additional features of our model and/or the use of data-consistent parameter values did not support mutualistic coexistence in any environmental condition we tested (however, competitive coexistence is possible). The discrepancy suggests that our model and even the classical resource-service exchange theory have missed some qualitative or quantitative details that are the key to understanding of syntrophic mechanisms in this specific example.

What could we have missed? Since mutualism occurs when the reciprocal benefits associated with cross-feeding outweigh competitive costs[46], our model should logically predict either or both of lower benefits and higher costs than the null expectation from simpler models. In the classical theory of syntropy, it is typically assumed that leaking chemicals are by-products which are inhibitory to producers but beneficial to consumers[45]. Since acetate was shown to inhibit growth of both cell types (Fig. 2D) and acetate specialist (the consumer) is more sensitive, its population density may be insufficient to reward the glucose specialist to a level that allows benefits higher than costs. On the other hand, costs are potentially similarly high since both cell types are polymorphic and share similar glucose uptake kinetics. We estimated that the relative difference in their maximum growth rates is 12%, which is much smaller than the observed value in experiments (33%)[16]. This quantitative difference may be important considering that the competition is stronger between populations with similar nutrient acquisition strategies. Recently, it was theoretically proposed that controlled metabolic leakages optimize resource allocation and can be beneficial to producers even under nutrient limitation[47]. We speculate that in case where acetate overflow improves, rather than negatively impacts, the growth of producers, the likelihood of forming a mutualistic pair between two cell types would be much higher. Overall, the cost-benefit nature of the cross-feeding interaction between polymorphic *E. coli* strains is more complex than thought before and warrants further research.

So far, the current framework has been applied to well-characterized communities with known chemicals and associated interactions. Can it be applied to infer community structure of complex microbiomes (e.g., human gut microbiome) where most of the metabolic exchanges involved in microbe-microbe interactions are still unknown? Our model has the potential if some technical challenges can be solved. First, direct modeling of a real-world microbiome with hundreds of species would be hurdled by too many unknown kinetic parameters. One way to solve this problem is to simply ignore the rare species[35]. Another—arguably better—approach might be by grouping species composition into functional guilds using unsupervised methods that infer those groups from the data alone[48], or to use prior knowledge from genomics or taxonomy to create such functional groups. Second, inferring chemical mediators within a community of interacting populations is a nontrivial task. It can be facilitated by prior knowledge such as searching the literature or leveraging systems biology tools such as community-level metabolic network reconstruction[49]. Finally, our model is nonlinear, so that an efficient and robust nonlinear regression approach for parameter estimation is essential. Manual parameter selection is often the only possible approach for small datasets like the experimental systems we analyzed here. Indeed, non-linear optimization algorithms often fail to converge to a realistic set of parameters. Although we chose the manual method to calibrate our models in this proof-of-concept study, manual fitting requires an expert operator and is a time-consuming process, which for now precludes it from being applied to large-scale microbial communities. On the positive side, the process of trial-and-error was greatly improved by the speed at which the intermediate-scale model runs simulations on a regular desktop computer. Beyond these technical issues, the model itself can be extended in multiple ways such as incorporating mechanisms of resource allocation and non-metabolite-mediated interactions and, despite any present limitations, we anticipate that network inference using mechanism-explicit models can open new avenues for microbiome research towards more quantitative, mechanistic, and predictive science.

## Methods

**General.** The modelling framework was developed by integrating a classical ecology model for population and nutrient dynamics and a coarse-grained description of cell metabolism. Custom MATLAB (The MathWorks, Inc., Natick, MA, USA) codes were developed to perform computational simulations and analyses of all three cross-feeding communities. Parameter values were obtained from either literature or a combination of manual and automatic data fitting. See Supplementary Information for a detailed description of the general modeling framework, the specific models for each of the three communities, as well as their theoretical analyses.

**Simulation.** Deterministic trajectories and their steady states in batch and chemostat conditions were simulated by solving the differential equations from the beginning to the end. Simulations of serial dilution transfer were slightly different in the aspect that the equations were only integrated within each day. The initial condition at the beginning of a day was obtained by dividing all population densities and nutrient concentrations at the end of the previous day by the dilution factor and resetting the feed medium glucose concentration to its initial value at day 0.

**Network perturbation.** External perturbations were exerted upon the steady state of the 4-auxotroph community. Nutrient downshift was simulated by decreasing the feed medium concentration of glucose at time 0. The effects of antibiotics targeting amino acid auxotroph $i$ was simulated by multiplying the growth rate of the auxotroph by an inhibitory term, i.e., $J_i^{grow} \rightarrow J_i^{grow}/(1 + [A]/K_i)$, where $[A]$ is the antibiotic concentration and $K_i$ is the inhibition constant. We assumed antibiotic concentration remains constant and chose $K_i = 1\ \mu M$. The cheaters of each amino acid auxotroph were simulated by turning off all amino acid leakages of the auxotroph. They were mixed with the resident community in varying ratios at the beginning of simulation. For all three perturbation types, the feed medium glucose concentration is 0.2 wt% in the unperturbed condition and serial dilution was run to steady state at 60 days.

## Acknowledgements

We thank Dr. Michael Mee, Dr. Harris Wang, and Dr. Jennifer Reed for provision and clarification of their experimental data. We also thank Dr. Jinyuan Yan for proofreading early drafts. This work was supported by NIH grants U01 AI124275 and R01 AI137269-01 to J.B.X. The funders had no role in study design, data collection and analysis, decision to publish, or preparation of the manuscript.

## Data availability

The simulation data that support the conclusions of this study are available from authors upon reasonable request.

## Code availability

The source codes for simulations of the three cross-feeding communities are available from https://github.com/liaochen1988/coarse-grained-ecology-models-for-microbial-community.

**Figure Legends:**

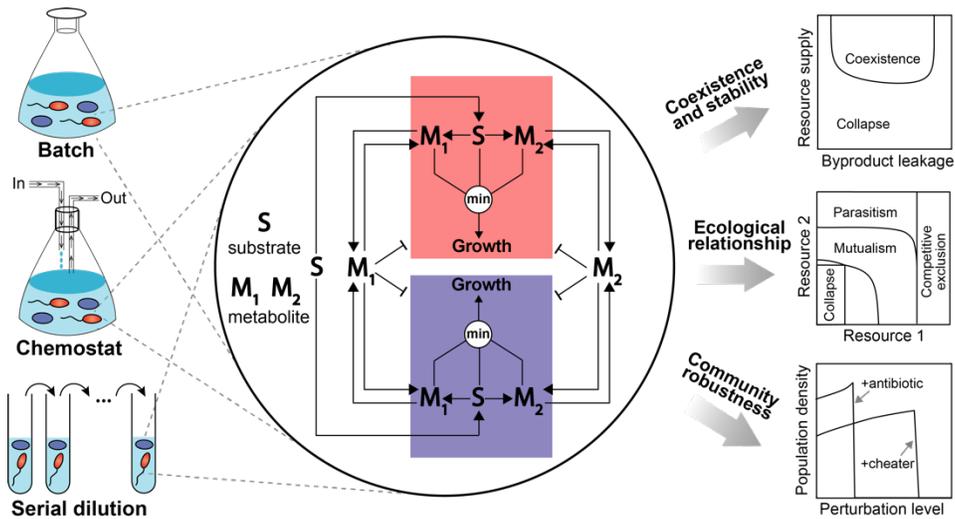

**Figure 1 | Schematic diagram illustrating our model and its potential applications in microbial ecology research.** A distinguishing feature of our microbial community model is that each community member harbors a coarse-grained metabolic network. Briefly, the metabolic network transforms substrates (**S**) to byproduct metabolites (**$M_1$, $M_2$**) and then to biomass whose production rate is set by the supply flux of the most limiting resource among all substrates and metabolites. For simplicity, the network is visually illustrated using one substrate and two metabolites but it can be extended to any number of molecules. Enabled by the simplified metabolic network, different community members can interact through a variety of mechanisms, including exploitative competitions for shared substrates, cooperative exchanges of nutritional metabolites, and direct inhibition by secreting toxic metabolites. Using training data from batch, chemostat or serial dilution cultures, our model can be parameterized to infer microbial processes underlying the data and then used to explore ecological questions and generate testable predictions. Pointed arrows denote the material flow and blunt-end arrows represent growth inhibition.

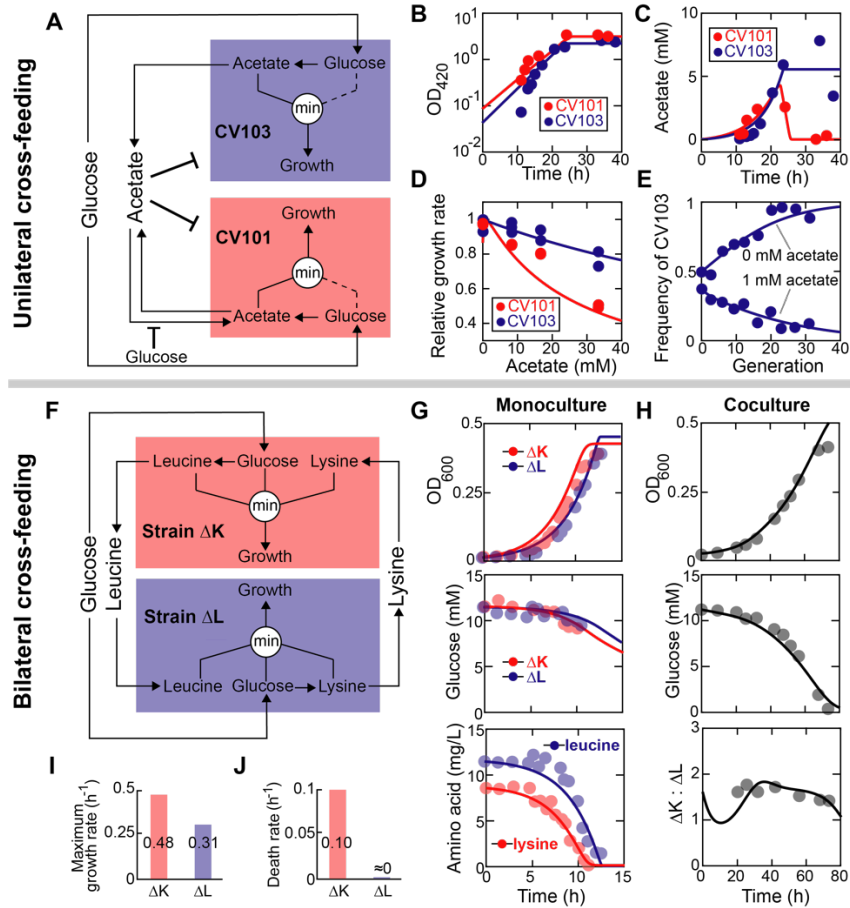

**Figure 2 | Model validation using two simple cross-feeding ecosystems.** (A-E) Unilateral acetate-mediated cross-feeding. (A) Schematic diagram of the model. The glucose specialist (CV103) and acetate specialist (CV101) are two *E. coli* mutants with different metabolic strategies[16]: the glucose specialist has improved glucose uptake kinetics while the acetate specialist is able to use acetate as an additional carbon source. At high concentrations the acetate inhibits the growth of both strains and its uptake by the acetate specialist strain is weakly repressed by the glucose. We assume that glucose and acetate are fully substitutable resources and simplify the model by limiting bacterial growth dependence to acetate alone (indicated by dashed lines; see experimental support of this hypothesis in Supplementary Texts 1.2.1). (B-E) Manual model calibration. Circles: experimental data; lines: simulations. (B,C) 0.1% glucose-limited batch monoculture without supplementing acetate[16]. (D) 0.0125% glucose-limited batch monoculture

supplemented with different concentrations of acetate[50]. (E) 0.00625% glucose-limited chemostat (dilution rate: 0.2 h$^{-1}$) coculture with (1 mM) and without acetate supplementation[16]. (F-J) Bilateral amino-acid-mediated cross-feeding. (F) Schematic diagram of the model. The *E. coli* lysine auxotroph (ΔK) and leucine auxotroph (ΔL) compete for glucose while additionally acquiring essential amino acids from each other. Growth of each auxotroph is determined by the more limiting resource between glucose and the amino acid it needs to grow. (G,H) Manual model calibration. Circles: data; lines: simulation. (G) 2 g/L glucose-limited batch monoculture supplemented with 10 mg/L amino acids[17]. (H) 2 g/L glucose-limited batch coculture without amino acid supplementation. (I,J) Inferred maximum growth rate when all limiting nutrients are supplied in excess (I) and death rate (J) of ΔK and ΔL strains.

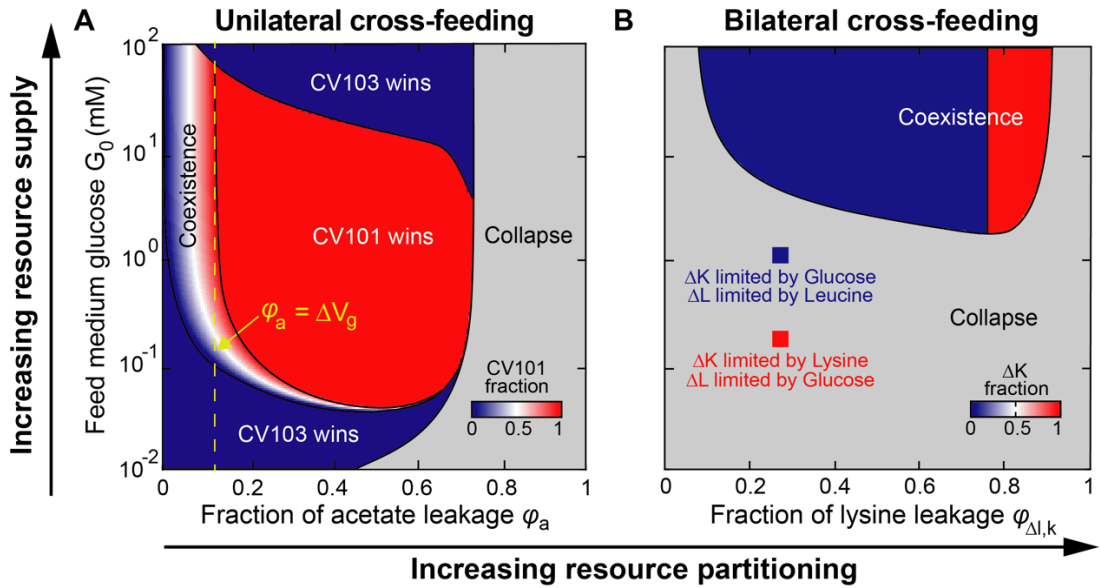

**Figure 3 | Impacts of resource supply and partitioning on coexistence of cross-feeders.** Steady state compositions of the unilateral (A) and the bilateral (B) cross-feeding communities are shown for varied levels of resource supply and partitioning. In (A), $\Delta V_g$ represents the relative difference in maximum glucose uptake rates between the glucose and acetate specialists, and gives the theoretical threshold of acetate leakage fraction above which the region of coexistence shrinks substantially. In (B), the leucine leakage fraction $\varphi_{\Delta k,l}$ was fixed at 0.5 and the lysine leakage fraction $\varphi_{\Delta l,k}$ was varied. Supplementary Fig. 2 shows that the symmetric choice that fixes $\varphi_{\Delta l,k}$ and varies $\varphi_{\Delta k,l}$ does not change the pattern of coexistence. All chemostat simulations were run at the dilution rate of 0.1 h$^{-1}$. CV103: glucose specialist; CV101: acetate specialist; ΔK: lysine auxotroph; ΔL: leucine auxotroph.

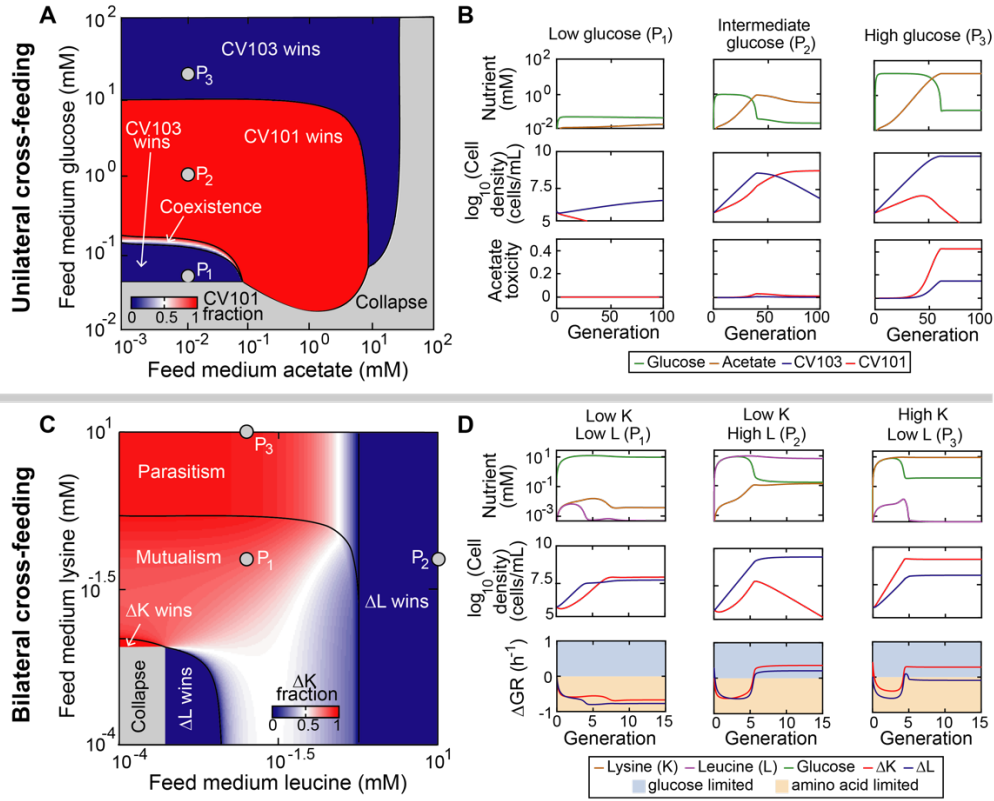

**Figure 4 | Impacts of nutrient supplementation on ecological relationships between cross-feeders.** Steady state compositions (A,C) and representative system dynamic trajectories (B,D) of the unilateral (A,B) and the bilateral (C,D) cross-feeding communities are shown for different levels of nutrient supplementation. In (B), acetate toxicity was defined as the ratio of growth rates between the presence and the absence of acetate. In (D), ΔGR was defined as the growth rate difference between amino-acid-limiting and glucose-limiting conditions. A positive or negative value of ΔGR indicates that cell growth is limited by glucose or amino acid respectively. The dilution rates used to run chemostat simulations of the unilateral and bilateral cross-feeding communities are 0.2 and 0.1 $h^{-1}$ respectively. CV103: glucose specialist; CV101: acetate specialist; ΔK: lysine auxotroph; ΔL: leucine auxotroph.

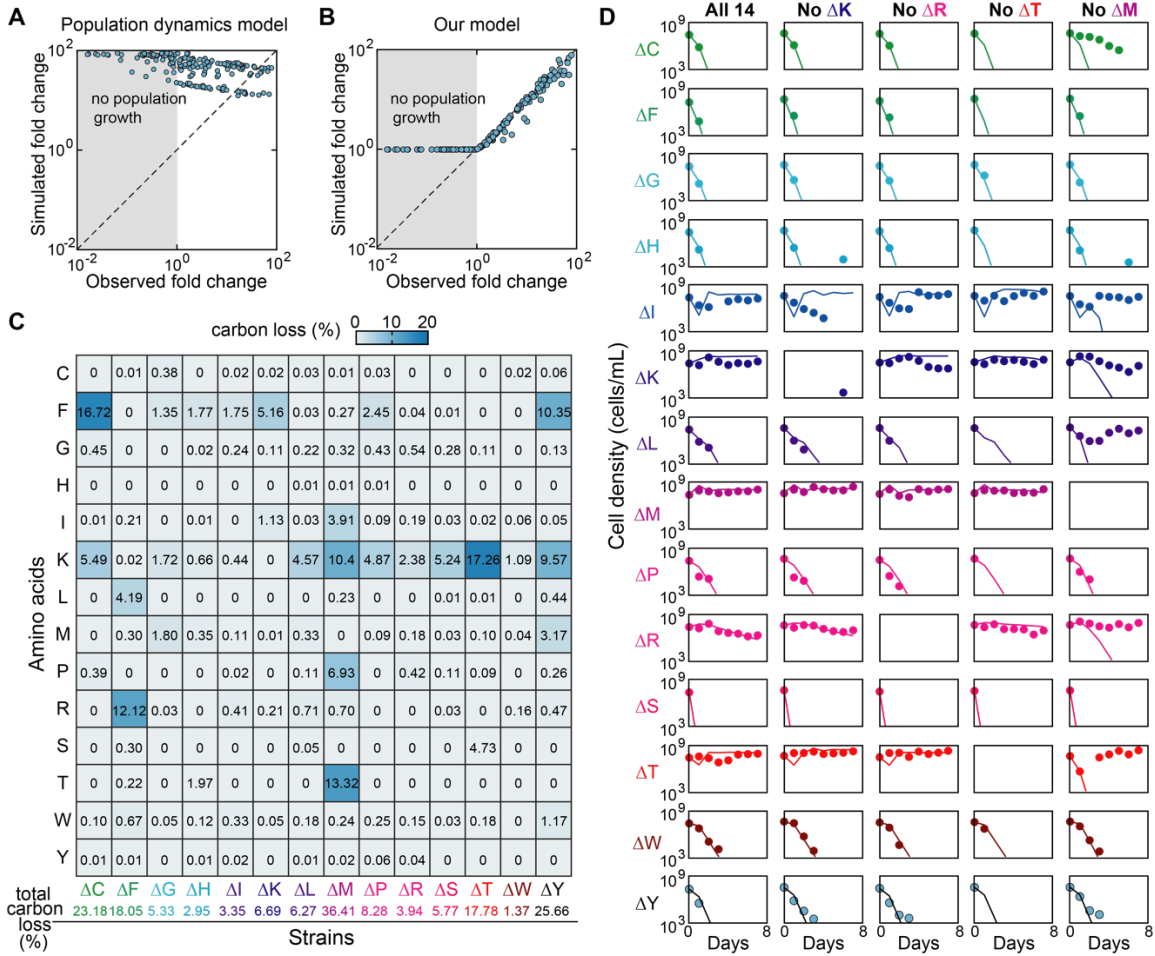

**Figure 5 | Modeling a consortium of 14 amino acid auxotrophs.** (A,B) Comparison of fold changes in observed[18] and simulated cell densities in batch coculture of all possible pairwise combinations of 14 *E. coli* amino acid auxotrophs. The population dynamics model and its associated parameters were adopted from Mee *et al.*[18]. (C) Predicted amino acid leakage profiles for the 14 auxotrophs. Each value in the matrix describes the fraction of carbon loss due to release of the amino acid in the row by the auxotroph in the column. (D) Comparison of the observed[18] (circles) and the simulated (lines) population dynamics in 7-day 100-fold serial dilution of one 14-auxotroph and four 13-auxotroph communities. Abbreviations: cysteine auxotroph (ΔC), phenylalanine auxotroph (ΔF), glycine auxotroph (ΔG), histidine auxotroph (ΔH), isoleucine auxotroph (ΔI), lysine auxotroph (ΔK), leucine auxotroph (ΔL), methionine auxotroph (ΔM),

proline auxotroph (ΔP), arginine auxotroph (ΔR), serine auxotroph (ΔS), threonine auxotroph (ΔT), tryptophan auxotroph (ΔW), and tyrosine auxotroph (ΔY).

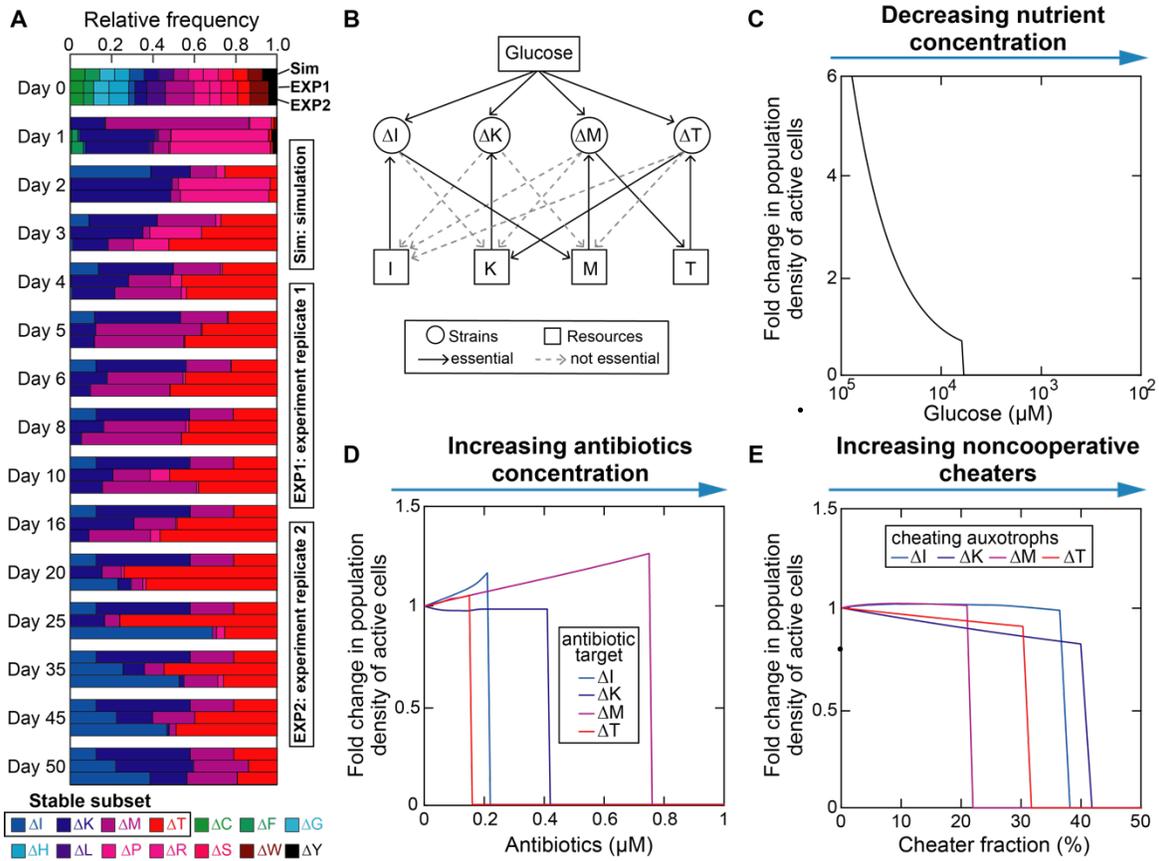

**Figure 6 | Collapse of mutualistic cross-feeding network following external perturbations.** (A) Emergence of stable coexistence of a four-auxotroph subset (ΔI, ΔK, ΔM, ΔT) over 50 daily passages. The two replicates of experimental observations were adopted from Mee *et al.*[18]. We used the same simulation parameters as in Fig. 5D except for a longer simulation time. See Fig. 5 legend for abbreviations of the names of amino acid auxotrophs. (B) Predicted bipartite interaction network of the subset. The network contains resource nodes (I, K, M, T for isoleucine, lysine, methionine, and threonine respectively) and consumer nodes (ΔI, ΔK, ΔM, ΔT are their corresponding auxotrophs), and each directed link describes a resource-consumer relationship. (C-E) External perturbations, including decreasing nutrient concentration (C), increasing antibiotic concentration (D), and introducing noncooperative cheaters (E), result in an abrupt collapse of the community when the perturbation level exceeds a certain threshold.